\documentclass[12pt]{iopart} 
\usepackage{epsfig,multicol,iopams,graphicx}

\def \M {\mathcal{M}} 
\def \J {\mathcal{J}} 
\def \v {\nu} 
\def \w {\omega}
\def \II {\mathcal {I}}
\def \I {\mathfrak {I}}
\def \pre{Preprint}

\begin{document}

\title{Area spectra versus entropy spectra of black holes in topologically massive gravity}

\author{Yongjoon Kwon and Soonkeon Nam}

\address{ Department of Physics and Research Institute of Basic Sciences, \\
 Kyung Hee University, Seoul 130-701, Korea}

\eads{\mailto{emwave@khu.ac.kr}, \mailto{nam@khu.ac.kr}}



\begin{abstract}
The area and entropy spectra of black holes in topologically massive gravity with the gravitational Chern-Simons term are calculated. The examples we consider are the BTZ black hole and the warped ${\rm AdS}_3$ black hole.  
For the non-rotating BTZ black hole we find that the gravitational Chern-Simons term does not affect the area and entropy spectra, so that they are equally spaced. For the rotating BTZ black hole, the spectra of the inner and outer horizon areas are not equally spaced and dependent on the coupling constant $\v$ of the gravitational Chern-Simons term. However the entropy spectrum is equally spaced and independent of the coupling constant $\v$. For the warped ${\rm AdS}_3$ black hole we find again that the entropy spectrum is equally spaced and independent of  the coupling constant $\v$, while the spectra of the inner and outer horizon areas are not equally spaced and dependent on  the coupling constant $\v$. Our result implies that the entropy spectrum has a universal behavior regardless of the presence of the gravitational Chern-Simons term, and therefore the entropy is more `fundamental' than the horizon area. We propose to use this universal behavior of entropy spectrum as criteria of a correct quasinormal mode.
\end{abstract}
%



\maketitle

\section{Introduction}

The quantum nature of black holes represents a significant and valuable property as a tool in understanding quantum gravity.
This can be seen from the quantization of black hole horizon area which follows from the fact that there can be  emission or absorption of quanta from a black hole. 
It was first proposed that the black hole horizon area is an adiabatic invariant and should be quantized by Ehrenfest principle, which says that any classical adiabatic invariant corresponds to a quantum entity with discrete spectrum \cite{be}. By considering the minimum change of horizon area in the process of absorption of a test particle it was proposed that the horizon area is linearly quantized as follows \cite{be}:
\begin{equation}
A_n =\gamma   n   \hbar ~~,~~ n=0,1,2,.. ,
\end{equation}
where $\gamma$ is an undetermined dimensionless constant. 
Since this proposal, the horizon area spectra of black holes have been investigated in various ways. 
Among them, it was realized that the quasinormal mode of a black hole which is the characteristic sound of the black hole would play a significant role in obtaining area spectrum of the black hole.  
Based on Bohr's correspondence principle, by considering the real part of the asymptotic quasinormal mode of a black hole as a transition frequency in the semiclassical limit, the dimensionless constant $\gamma$ could be determined  \cite{hod}. 
For the example of the Schwarzschild black hole,  the real part of the asymptotic quasinormal mode is given by  $ \w_R = {{ { \rm ln 3} } \over {8 \pi M}} $ \cite{nol} and  the dimensionless constant  $\gamma$  was determined as 
%
\begin{equation} \label{ho}
\gamma = 4  \, {\rm ln} 3 ~,
\end{equation}
where  the relations $A=16 \pi M^2$  and $\triangle M=\hbar  \w_R$ were used \cite{hod}.
%

Later, the area spectrum of the Schwarzschild black hole was reproduced in a different method, where  an adiabatic invariant of the system was used \cite{kun}.
It was noticed that given a system with energy $E$ and vibrational frequency $\w(E)$,  the quantity $I=  \int {dE / {\w(E)} } $ is an adiabatic invariant \cite{kun}. The real part of the quasinormal frequency was considered as the vibrational frequency and  Bohr-Sommerfeld quantization, $I=n \hbar$, was applied in the semiclassical limit. Since the change of the energy of a black hole is the change of the ADM mass $M$ of the black hole, the formula was obtained as follows:
\begin{equation} \label{kuf}
I = \int {dE \over {\w_R} } = \int {dM \over {\w_R} }  
=n \hbar ~.
\end{equation}
%
From this formula,  it was obtained that the area spectrum of the Schwarzschild black hole is the same as the previous result (\ref{ho}) and  equally spaced. 

Following the formula (\ref{kuf}), a modified adiabatic invariant $\II$ was considered  for a rotating black hole.
The modified adiabatic invariant $\II$  was given by \cite{setvag}
\begin{equation} \label{mkuf}
\II= \int {{dM -\Omega   dJ}  \over {\w_R} } ~,
\end{equation}
where $\Omega$ is the angular velocity at horizon. This adiabatic invariant $\II$ was quantized  as $\II =n \hbar$ via Bohr-Sommerfeld quantization. 
This was applied for the Kerr black hole and the area spectrum was not equally spaced \cite{setvag}. 

Recently, it was proposed that since a perturbed black hole has to be described as a damped harmonic oscillator, the characteristic classical frequency $\w_c$ should be identified with the transition frequency between quantum levels $(\w_{0})_k$, which is defined as $(\w_0)_k = ( {\sqrt{ {\omega _R}^2+{\omega _I}^2 } })_k $, in the semiclassical limit \cite{magi}.
Since  most black holes have $\omega _{I} \gg \omega _{R}$ at highly excited levels, we have  $\omega _0 \sim \omega _I  $ rather than $ \omega _0 \sim \omega _R$. 
Therefore for the highly damped mode $( \omega _{I} \gg \omega _{R} )$ we have $\w_{c} = (\w_ {0})_{k} - (\w_ {0})_{k-1}  \simeq  ( | \w_ {I} | )_{k} - ( | \w_ {I} | )_{k-1}$, where $k \in \ N $ and $ k \gg 1$.
With this newly identified frequency, the area spectrum of the Schwarzschild black hole was obtained from the relation $\triangle M = \hbar \omega _c= { \hbar \over {4 M}}$ as follows \cite{magi}:
\begin{equation} \label{revi1}
A=8 \pi n \hbar ~.
\end{equation}
This area spectrum is equally spaced, but has different spacing from the previous result (\ref{ho}).
By using  the newly identified frequency $\omega _c$ instead of $\omega _R$ in the formula (\ref{kuf}), the  area spectrum (\ref{revi1}) was reproduced  \cite{med, wei}. 
%
%
With this transition frequency $\omega _c$ the area spectrum of the Kerr black hole was reconsidered by using  Bohr-Sommerfeld quantization of the modified adiabatic invariant (\ref{mkuf})  \cite{med, vage}. 
It was obtained again that the area spectrum of the Kerr black hole is not equally spaced, which is still in contradiction to Bekenstein's proposal \cite{be}.
In our recent work \cite{sk}, however, it was pointed out that Bohr-Sommerfeld quantization should be applied to an action variable of the system rather than an adiabatic invariant, and that not every adiabatic invariant is an action variable. It is well known that the action variable via Bohr-Sommerfeld quantization  should be given by
%
\begin{equation}
{\mathfrak I}= {1 \over {2 \pi}} \oint p dq  = n \hbar ~~,~~ (n=0,1,2.. )~.
\end{equation}
%
Based on Bohr's correspondence principle, the transition frequency $\w_c$ of a black hole  in the semiclassical limit is considered as the oscillation frequency in a classical system of periodic motion.
So, the action variable of the corresponding classical system is identified and quantized via Bohr-Sommerfeld quantization in the semiclassical limit as follows \cite{sk}:
\begin{eqnarray} \label{mf}
{\mathfrak I} = \int { dE \over {\omega _c }} = n \hbar ~~,~~ (n=0,1,2.. )~.
\end{eqnarray}
%
A black hole with transition frequency can be considered as a classical system of periodic motion with oscillation frequency equal to the transition frequency in the semiclassical limit. Therefore the formula (\ref{mf}) holds for a black hole with quasinormal mode regardless of whether it is rotating or not \cite{sk}.
As an example for a rotating black hole, the BTZ black hole was considered in our previous work \cite{sk} and by using the formula (\ref{mf})  we obtained that the area and entropy spectra are  equally spaced.
%

In this paper we would like to apply the method to topologically massive gravity theory in three dimensions  \cite{deser, deser2}, and investigate the area and entropy spectra of black hole solutions in that theory, where  the gravitational Chern-Simons term  is considered in the Einstein-Hilbert action with the cosmological constant. 
The reason we consider this is that  the simple Bekenstein-Hawking area law, $S=A/(4 G \hbar)$, is no longer true, due to the presence of the gravitational Chern-Simons term. 
We then face a small puzzle. Can either of the area spectrum or the entropy spectrum have a universal behavior of equally spaced spectrum? The physical quantity with the universal behavior should be regarded as more `fundamental'. So, it is a very interesting question to ask if the black holes in the topologically massive gravity have such a universal behavior and which one is  `fundamental'.
 If any of the spectra indeed has such a behavior, it will provide a substantial support for our earlier proposal  \cite{sk}. 
Further motivation for the inclusion of the gravitational Chern-Simons term comes from the fact that it rises naturally in string theory. Sometimes, the compactification of superstring theory down to odd dimensional  spacetimes gives rise to the gravitational Chern-Simons term. In particular for the three dimensional case, it becomes the topologically massive gravity.  As we know, in three dimensions there are no propagating degrees of freedom  even though asymptotically $AdS_3$ black hole solution exists \cite{BTZ}. However, the topologically massive gravity has a single massive graviton mode because of the gravitational Chern-Simons term \cite{deser}.
So,  we will consider black holes in three dimensional topologically massive gravity with a negative cosmological constant, whose action is given by \cite{deser, deser2, strom}
\begin{eqnarray} \label{action}
\fl {\rm I} &=& {\rm I_{EH} } + {\rm I_{CS} }  \nonumber \\
\fl &=& {1 \over {16 \pi G}} \int _M d^3 x { \sqrt{-g}  \left( R+{ 2 \over {l^2} } \right)} 
- {l \over {96 \pi G \v}} \int _M d^3 x  {\sqrt{-g}   \varepsilon ^{\lambda \mu \nu} \Gamma ^{r} _ {\lambda \sigma} \left( \partial_\mu \Gamma ^{\sigma} _{r \nu} +{2 \over 3}  \Gamma ^{\sigma} _{\mu \tau}  \Gamma ^{\tau} _{\nu r} \right)} ,
\end{eqnarray}
where $\v$ is the coupling constant and  $\varepsilon ^{\tau \sigma \mu}=1 / \sqrt{-g} $ is the Levi-Civita tensor. The first term $\rm I_{EH} $ is the Einstein-Hilbert action with a negative cosmological constant and the second term $\rm I_{CS} $  is the gravitational Chern-Simons term.
Varying this action with respect to the metric, we obtain the equation of motion;
\begin{equation} \label{revieq}
G_{\mu \nu} -{1 \over l^2}   g_{\mu \nu} + {l \over  {3 \v} }  C_{\mu \nu} =0 ~,
\end{equation}
where $G_{\mu \nu} $ is the Einstein tensor  and $C_{\mu \nu} $ is the Cotton tensor;
\begin{eqnarray}
G_{\mu \nu} =  R_{\mu \nu} -{1 \over 2} g_{\mu \nu}  R ~~, ~~ C_{\mu \nu} =  \varepsilon  _{\mu} ^{\alpha \beta}    \nabla _\alpha \left( R_{\beta \nu} -{1 \over 4}  g_{\beta \nu}  R \right) ~.
\end{eqnarray}
%
We will consider two examples of black holes as the solutions of the equation of motion (\ref{revieq}). 
One is the BTZ black hole and the other is the warped ${\rm AdS}_3$ black hole which is asymptotic to warped ${\rm AdS}_3$. The warped ${\rm AdS}_3$ black hole without naked closed timelike curves is the spacelike stretched black hole  for $\v >1$. 
The warped ${\rm AdS}_3$ black hole for $\v > 1$ has the non-vanishing Cotton tensor, $C_{\mu \nu} \neq 0$, while the BTZ black hole and the warped ${\rm AdS}_3$ black hole for $\v = 1$ have the vanishing Cotton tensor, $C_{\mu \nu}=0$. 
Indeed  the metrics of the BTZ black hole and the warped ${\rm AdS}_3$ black hole  for $\v = 1$ are related by a coordinate transformation each other \cite{strom}. 

We will apply the formula (\ref{mf}) for the BTZ black hole and the warped ${\rm AdS}_3$ black hole
and calculate  the area and entropy spectra of the black holes with the gravitational Chern-Simons term.
%
This paper is organized as follows:
First, we will consider the BTZ black hole with the Chern-Simons term. 
For the rotating case we  find that  the inner horizon area as well as the outer horizon area is quantized and the area spectra are not equally spaced and dependent on the coupling constant $\v$.
%
However the entropy spectrum is equally spaced and independent of the coupling constant $\v$. This entropy spectrum is the exactly same as for  the rotating BTZ black hole without the Chern-Simons term obtained in  \cite{sk}. 
This implies that the entropy is more `fundamental' than the horizon area.
For the non-rotating case  both area and entropy spectra are equally spaced and independent of the coupling constant $\v$. The area and entropy spectra are the exactly same as for the non-rotating BTZ black hole without the Chern-Simons term obtained in \cite{sk}. 
Therefore the Chern-Simons term does not affect the area and entropy spectra of the non-rotating BTZ black hole, different from the rotating case.
%

%
Next, we will consider  the warped ${\rm AdS}_3$ black holes for $\v \ge1$. 
%
%
For  $\v = 1$ case, we find that the spectra of the both inner and outer horizon areas are equally spaced and the entropy spectrum is also equally spaced. These area and entropy spectra are  exactly same as for the rotating BTZ black hole with the Chern-Simons term for $\v=1$, as it is expected. 
%
%
For $\v > 1$, we find that while the spectra of the inner and outer horizon areas are not  equally spaced and dependent on the coupling constant $\v$, the entropy spectrum is equally spaced and independent of the coupling constant $\v$.
This behavior of the area and entropy spectra is just like the rotating BTZ black hole case with the Chern-Simons term.
Moreover the entropy spectrum is exactly same as for the rotating BTZ black hole regardless of whether the Chern-Simons term is considered or not. 
It implies that the entropy spectrum has a universal behavior regardless of the presence of the gravitational Chern-Simons term. 

In the last section, we propose that the universality of the equally spaced entropy spectrum can be  used for criteria for checking if the calculated quasinormal modes of a black hole are correct.
For the quasinormal modes of the warped ${\rm AdS}_3$ black hole for $\v>1$, there are two different calculations  which give different results in  \cite{bin, bin2} and  \cite{0912}. 
The quasinormal modes  in \cite{bin, bin2} are obtained from the vanishing Dirichlet boundary condition at radial infinity, while in \cite{0912}  the quasinormal modes are obtained  without imposing boundary condition at radial infinity.  
Since the warped ${\rm AdS}_3$ black hole for $\v>1$ has the finite effective potential at radial infinity, it is not clear whether we can impose the vanishing Dirichlet boundary condition at radial infinity.  However, as we propose, from the universality of the entropy spectrum we can find which one is right.
In light of the criteria that the equally spaced entropy spectrum should be universal, we find that the quasinormal modes in  \cite{bin, bin2} are ruled out, since they give non-equally spaced entropy spectrum.
%

\section{Area and entropy spectra of the BTZ black hole with the gravitational Chern-Simons term}  

We will consider the rotating BTZ black hole with the gravitational Chern-Simons term. 
The metric of the rotating BTZ black hole  is given by \cite{BTZ}
\begin{eqnarray}
\fl ds^2= \left(8 G M - { r^2 \over l^2} -{ {16 G^2 J^2} \over {r^2} } \right) dt^2+ { {dr^2} \over {\left(-8 G M+{ r^2 \over l^2} +{ {16 G^2 J^2} \over {r^2} }\right)} }  + r^2 \left( d\phi -{{4 G J} \over {r^2} } dt \right) ^2 ,
\end{eqnarray}
where the cosmological constant is given by $\Lambda=-{1 \over l^2}$.
The mass $M$ and angular momentum $J$ of the black hole can be expressed in terms of the outer and inner horizons, $r_\pm$, as follows:
\begin{equation}
M= { {  {r_+ ^2}+ {r_- ^2} } \over {8 G l^2}} ~~,~~ J= { {  2 {r_+}  {r_-} } \over {8 G l}} ~.
\end{equation}
The quasinormal modes are same as for the rotating BTZ black hole without the Chern-Simons term, since the  metric  is unchanged even though the Chern-Simons term is considered.
For the rotating BTZ black hole the two families of the quasinormal modes  are given by \cite{bir}
\begin{eqnarray}
\w_{R} &=& -{\frac {m}{l}}- i {\frac {  (r_{+}+r_{-} ) \left( 2  k+1+{\sqrt {1+\mu} } 
 \right) }{{l}^{2}}} ~, \\
\w_{L} &=& {\frac {m}{l}}- i {\frac {  (r_{+}-r_{-} ) \left( 2  k+1+{\sqrt {1+\mu} } 
 \right) }{{l}^{2}}}  ~,
\end{eqnarray}
where  $m$ is  the angular quantum number and $ k$ is the overtone quantum number which comes from the boundary condition in the radial direction. The mass parameter $\mu$ is given by $\mu \equiv u^2 l^2 / {\hbar ^2}$, where $u$ is the mass of the scalar field.

At large $k$ for a fixed $\vert m \vert$, in particular for $k \gg \vert m \vert$, the two transition frequencies corresponding to each quasinormal mode are obtained as follows:
\begin{eqnarray} \label{btzq}
\w_{R c} &=&  {{2  (r_{+}+r_{-} ) } \over {l^2}} = { {2 \sqrt{8 G M+{8 G J / l}} } \over  l} ~,\\
\label{btzq1}
\w_{L c} &=&  {{2  (r_{+}-r_{-} ) } \over {l^2}} = { {2 \sqrt{8 G M-{8 G J /l}} } \over  l}  ~.
\end{eqnarray}
We consider the two action variables, $\I_R$ and $\I_L$, corresponding to each possible transition frequency.
From the formula (\ref{mf}) two action variables are obtained and quantized.
We should note that when the Chern-Simons term is considered, the change of the energy of the black hole  is not  the  change of  the ADM mass $M$  of the black hole. Because of the Chern-Simons term, the conserved charge $\M$ of the BTZ black hole, which is called ADT mass, is given by \cite{clem, sol}
\begin{equation} \label{sole}
\M= M+{ J \over {3 \v l} } ~,
\end{equation} 
where $M$ and $J$ are the mass and the angular momentum for the BTZ black hole in the absence of the Chern-Simons term.
So, the change of the energy of the black hole should be considered as  the change of the ADT mass $\M$, i.e. $dE=d\M  $.
Note that (\ref{sole}) holds only for the stability bound, $\v \ge 1/3$, which is from the requirement that central charges should be non-negative \cite{sol, song}.
Therefore we consider only  the cases for  $\v \ge 1/3  $, and the action variable $\I$ is given by
\begin{equation} \label{btbt}
\I = \int { dE \over {\w_c} } = \int { d{\M} \over {\w_c} } =\int { dM \over {\w_c} } + {1 \over {3 \v l} } \int { dJ \over {\w_c} } ~.
\end{equation}
The two action variables, $\I_R$ and $\I_L$, are obtained and quantized as follows:
\begin{eqnarray} \label{gt11}
\fl  \I_{R} &=& \int { dM \over {\w_{Rc}} } + {1 \over {3 \v l} } \int { dJ \over {\w_{Rc}} } = {{(3 \v+1) l} \over {24 G \v}}   \sqrt{ 8 G M+8 G J/l} = n_R   \hbar ~,\\
\label{gt12}
\fl  \I_{L} &=&\int { dM \over {\w_{Lc}} } + {1 \over {3 \v l} } \int { dJ \over {\w_{Lc}} } = { {(3 \v-1) l} \over {24 G \v}} \sqrt{8 G  M-8 G J/l} = n_L   \hbar  ~.
\end{eqnarray}
First we will consider  the case for $\v > 1/3$.  
From (\ref{gt11}) we find that the total horizon area spectrum is given by
\begin{equation}
A_{tot} \equiv A_{out}+A_ {in} = 2  \pi  l   \sqrt{8 G  M+8 G J/l}  ={{48  \pi  G \v} \over {(3 \v+1)}}  n_R   \hbar  ~,
\end{equation}
where $n_R=0,1,2,.. $.
This spectrum is equally spaced and the spacing is dependent on the coupling constant $\v$.
From (\ref{gt12}) we find that the difference between the two horizon areas  is also equally spaced and dependent on the coupling constant $\v$ as follows:
\begin{equation} 
A_{sub} \equiv A_{out}-A_ {in} = 2  \pi  l   \sqrt{ 8 G M-8 G J/l} = {{48  \pi  G \v} \over {(3 \v-1)}}  n_L   \hbar   ~,
\end{equation}
where $n_L=0,1,2,.. $.
Therefore  the spectra of the outer and inner horizon areas are obtained as follows:
\begin{eqnarray} \label{gC_4}
A_{out} &=& { {24 G \pi \v} \over {9 \v^2 -1} }   \Big[   {(3 \v-1) n_R + (3 \v+1)   n_L} \Big]  \hbar ~,  \\
 \label{gC_5}
A_{in} &=& { {24 G \pi \v} \over {9 \v^2 -1} }   \Big[   {(3 \v-1)   n_R -(3 \v+1) n_L} \Big]  \hbar ~,
\end{eqnarray}
where $ (3 \v-1) n_R  \ge (3 \v+1)  n_L  $.
The area spectra are not equally spaced and dependent on the coupling constant $\v$, so that they are affected by the Chern-Simons term. 
Note that when we take the limit $\v$ goes to infinity, which means the vanishing of the Chern-Simons term in the action (\ref{action}), the area spectra become the ones for the rotating BTZ black hole without the Chern-Simons term obtained in \cite{sk} where the units, $c=8 G=1$, are used.

Now  let us find the entropy spectrum. When the Chern-Simons term is considered, the entropy of the BTZ black hole is not satisfied with the Bekenstein-Hawking area law, $S= {A \over {4 G \hbar}}$, any more. It  has another term  related to the inner horizon area as follows \cite{sol}:
\begin{equation} \label{gC_1}
S={1  \over {4 G \hbar} }   \left( { { 2   \pi   r_{+}  + {{ 2   \pi   r_{-}} \over {3 \v}}  }}   \right)= { 1 \over {4 G \hbar } }  \left( {{ A_{out}+{1\over {3 \v}}  A_{in} }  } \right) ~,
\end{equation}
where $r_{\pm}$ are the outer and inner horizons, respectively.
Using  (\ref{gC_4}) and (\ref{gC_5}), we find the equally spaced entropy spectrum as follows:
\begin{equation}  \label{gC_2}
 S= 2 \pi  (n_R + n_L) ~.
\end{equation} 
Therefore the spacing of the entropy spectrum is given by
 \begin{equation}  \label{gC_3}
\triangle S= 2 \pi ~.
\end{equation}
The entropy spectrum (\ref{gC_2}) is independent of the coupling constant $\v$ and exactly same as for the rotating BTZ black hole without the Chern-Simons term obtained in  \cite{sk}. 
Therefore the Chern-Simons term does not affect the entropy spectrum of the BTZ black hole, even though it affects the area spectra.  This implies that the entropy is more fundamental than the horizon area.

Next, the case for $\v=1/3$ is considered, which corresponds to chiral gravity.  
From (\ref{gt11}) and (\ref{gt12}),  we have only one non-vanishing action variable $\I_R  $; 
\begin{equation} \label{e5}
\I_R ={l \over {4 G}}   \sqrt{ 8 G M+8 G J/l} = n_R   \hbar ~.
\end{equation}
This quantization gives the equally spaced spectrum of the total horizon area;
\begin{equation} \label{ee55}
A_{tot} \equiv A_{out}+A_ {in} = 2  \pi  l   \sqrt{8 G  M+8 G J/l} =8  \pi  G n_R   \hbar ~,
\end{equation}
where $n_R=0,1,2,.. $.
However, we cannot obtain the spectrum of the difference between two horizon areas since  $\I_L=0$.
Nevertheless the entropy spectrum can be obtained, since the entropy  (\ref{gC_1}) for $\v=1/3$ is only proportional to the total horizon area. That is, when  $\v=1/3$, the entropy reduces to the following simple form:
\begin{equation}  
 S=  { { A_{out}+A_{in} } \over {4 G \hbar }  } =  { { A_{tot} } \over {4 G \hbar }  } ~.
\end{equation} 
Therefore using (\ref{ee55}), we find the equally spaced entropy spectrum as follows:
\begin{equation}  \label{e6}
 S= 2 \pi  n_R ~.
\end{equation} 
This entropy spectrum has the same spacing as for the rotating BTZ black hole for $\v > 1/3$, i.e. $\triangle S= 2 \pi $.

Lastly we will consider the non-rotating BTZ case whose horizon is located at $r_H = l  \sqrt{8 G M}$.
The quasinormal modes of the non-rotating BTZ black hole are given by  \cite{bir, cardo}
\begin{equation} \label{qn}
\omega =\pm {\frac {m}{l}} - i {\frac {2 \sqrt {8 G M} \left( k+1 \right) }{l}}  ~~,~~ (m \in Z  ,~ k =0,1,2,..)~,
\end{equation}
where $m$ is the angular quantum number and $k$ is the overtone quantum number of the quasinormal modes.
At large $k$ for a fixed $\vert m \vert$, in particular for $k \gg \vert m \vert$, we have a transition frequency in the semiclassical limit as follows:
\begin{equation}  
\w_c =  { {2 \sqrt{8 G M} } \over l} ={{ 2 r_H} \over l^2}  ~.
\end{equation} 
As mentioned before, when the Chern-Simons term is considered, the change of the energy of the black hole should be considered as the  change of  the ADT mass $\M$  which is given by (\ref{sole}).
However since the angular momentum $J$ is zero for the non-rotating BTZ black hole, we can consider the change of the energy of the black hole as the change of the ADM mass $M$,  $dE=dM$.
Therefore from  (\ref{mf}) the action variable $\I$ is calculated and quantized as follows:
\begin{equation}   \label{e2}
\I = \int { {dE} \over {\w_c} }  = \int { {dM} \over {\w_c} } = { l \over 2}   \int { {dM} \over {\sqrt{8 G M} }  } = { {l \sqrt{8 G M} } \over {8 G } } = n  \hbar ~.
\end{equation} 
From this quantization we find the  equally spaced area spectrum as follows:
\begin{equation} \label{e3}
A  = 2 \pi {r_H} =2  \pi  l   \sqrt{8 G  M} =16 \pi G n \hbar ~.
\end{equation}
Note that this spectrum does not depend on the  coupling constant $\v$. 
The area spectrum is the exactly same as for the non-rotating BTZ black hole without the Chern-Simons term obtained in  \cite{sk} where the units, $c=8 G=1$, are used.

Let us calculate the entropy spectrum.
When the Chern-Simons term is considered, the entropy  is given by (\ref{gC_1}).
For the non-rotating BTZ black hole, however, since the inner horizon area is zero, the entropy is simply given by 
\begin{equation}
S={ { 2   \pi   r_{+} }  \over {4 G \hbar} } = { {A} \over {4 G \hbar } } ~.
\end{equation}
It satisfies the Bekenstein-Hawking area law.
Note that the entropy  is also not affected by the Chern-Simons term.
Using the result of the area spectrum (\ref{e3}), we obtain  the equally spaced entropy spectrum as follows:
\begin{equation} \label{e4}
S=4 \pi n ~.
\end{equation}
This spectrum is also exactly same as for the non-rotating BTZ black hole without the Chern-Simons term obtained in  \cite{sk}. 
Therefore we conclude that for the non-rotating BTZ black hole the Chern-Simons term does not affect the area and entropy spectra.

\section{Area and entropy spectra of the warped ${\rm AdS}_3$ black hole}  

In this section, we would like to consider  the warped ${\rm AdS}_3$ black holes for $\v \ge 1$. The case $\v=1$ is related to the BTZ black hole considered in the previous section. The metric is given by  \cite{strom, bouch}
\begin{eqnarray} \label{wmt}
\fl {ds^2 } = l^2   \Biggl[ dt^2 + { {dr^2} \over { (\v^2 +3) (r- {r_+}) (r-{r_-}) }}+\left( 2 \v r-\sqrt{(\v^2 +3) {r_+}  {r_-}
}  \right) dt   d\theta  \nonumber \\
+ {r \over 4} \left(  3  (\v^2 -1) r+(\v^2 +3) ({r_+}+{r_-}) - 4 \v  \sqrt{ (\v^2 +3) {r_+}  {r_-} }   \right) d\theta ^2   \Biggr] ~,
\end{eqnarray}
where  $r_+$ and $r_-$ are the outer and inner horizons, respectively. 
Note that the coordinates $t$ and $r$  are dimensionless. 
It is another black hole solution of gravity theory with the gravitational Chern-Simons term.
The entropy of the warped ${\rm AdS}_3$ black hole is given by \cite{strom, bouch}
\begin{eqnarray} \label{went}
\fl S =  { {\pi  l} \over  {24 G \v \hbar} }      \left[ \left( 9 {\v}^{2} +3 \right) {r_+}  -  \left( {\v}^{2}+3 \right) {r_-} -4 \v\sqrt { (\v^2 +3) {r_+}  {r_-} }  \right] ~.
\end{eqnarray}
The conserved charges such as the ADT mass $\M$ and angular momentum $\J$ are given by \cite{strom, bouch}
\begin{eqnarray} \label{u1}
\fl \M &=& { { (\v^2 +3) } \over {24 G} } \left( { {r_+} + {r_-} - {1 \over \v} \sqrt{ (\v^2 +3) {r_+}  {r_-} } } \right ) ~, \\
\label{u2}
\fl \J &=& { { (\v^2 +3)  \v l } \over { 96 G} } \left [ \left(  {r_+} + {r_-} - {1 \over \v} \sqrt{(\v^2 +3) {r_+}  {r_-}}  \right )^2 -{ {(5 \v^2 +3) } \over {4 \v^2}}  ({r_+} - {r_-})^2  \right ] ,
\end{eqnarray}
which satisfy the first law of the black hole thermodynamics, $d\M ={\tilde T} dS+ {\tilde \Omega}  d\J$,  with the Hawking temperature ${\tilde T}$ and the angular velocity of the horizon $ {\tilde \Omega}$  given by \cite{strom}
\begin{eqnarray}
{\tilde T} &\equiv & {T \over l} = { {(\v^2 +3)} \over  {4 \pi l} } { {(r_+ -r_-)} \over {(2 \v r_+  -\sqrt{(\v^2 +3) r_+ r_- } ) }} ~,\\
{\tilde \Omega} &\equiv & {\Omega \over l}  = -{ { 2} \over { (2 \v r_+ -\sqrt{(\v^2 +3) r_+ r_-} )  l}} ~.
\end{eqnarray}
From the metric (\ref{wmt}) the outer and inner horizon areas are given by
\begin{eqnarray} \label{outer}
A_{out} &=& \pi l   \left(   2 \v  {r_+} -\sqrt{(\v^2 +3) {r_+} {r_-} }   \right)   ~,\\
 \label{inner}
A_{in} &=& \pi l    \left|    2 \v {r_-} - \sqrt{ (\v^2 +3)  {r_+} {r_-} }     \right| ~.
\end{eqnarray}
We are considering the non-extremal warped ${\rm AdS}_3$ black hole with the two horizons, $ r_+>r_-> 0$. 
The expression of the inner horizon area is considered for the following two cases:
\begin{eqnarray*} 
\fl  &&{\rm (i)} ~~ {{r_+} \over {r_-}}   >   { {4 \v^2} \over {(\v^2 +3)} } ~~{\rm case}  ~:  A_{in} =  \pi  l     \left( \sqrt{(\v^2 +3) {r_+} {r_-} } -2 \v {r_-}   \right) ~.\\
\fl  &&{\rm (ii)} ~~ 1< {{r_+} \over {r_-}}  <  { {4 \v^2} \over {(\v^2 +3)} }~~ {\rm case} ~:  A_{in} =  \pi  l   \left( 2 \v {r_-}  -\sqrt{(\v^2 +3) {r_+} {r_-} }   \right)~.
\end{eqnarray*} 
Note that the expression of the inner horizon area for $\v=1$  only corresponds to the case (i) and the expression of the inner horizon area for $\v>1$ is divided to the two cases as the above.

For the case (i), the total horizon area, $A_{tot} \equiv  A_{out}+A_ {in} $, and the area difference, $A_{sub} \equiv   A_{out}-A_ {in}$, are rewritten in terms of ADT mass $\M$ and angular momentum $\J$,  using  (\ref{u1}) and (\ref{u2});
\begin{eqnarray} \label{add}
\fl A_{tot} &=& 2  \pi   \v  l  ({r_+} - {r_-})  = {{16 \pi  \v l} \over { {\v}^{2}+3 }} \sqrt { {{ {6 G \v \left( 6 G {\M}^{2} \v -  \left( {\v}^{2}+3  \right) \J /l \right) }}  \over {{5 {\v}^{2}+3}}} } ~,\\
 \label{add2}
\fl A_{sub} &=&  2  \pi  \v  l  \left( {r_+} + {r_-} - \sqrt{ (\v^2+3)  {r_+} {r_-} }   \right) = { {48 \pi   G   \v  l} \over {\v^2 +3}}   \M ~.
\end{eqnarray}
For the case (ii) it is easily found that  $A_{tot}$ and $A_{sub}$ are interchanged each other.

Now let us calculate the area and entropy spectra  from the quasinormal modes of the warped ${\rm AdS}_3$ black hole by using the formula (\ref{mf}).
Usually the quasinormal modes can be obtained from the radial boundary conditions of the wavefunction, which is a solution of the wave equation obtained by a perturbation. In particular for black holes in AdS background, the vanishing Dirichlet boundary condition  at radial  infinity is imposed, since the effective potential in the wave equation  is divergent at radial infinity \cite{cardo, horo}.
In \cite{bin, bin2} the quasinormal modes of the warped ${\rm AdS}_3$ black hole are obtained from the vanishing Dirichlet  boundary condition. 
Since the warped ${\rm AdS}_3$ black hole for $\v=1$  has the divergent effective potential at radial infinity, the appropriate boundary condition which should be imposed at radial infinity is the vanishing Dirichlet  boundary condition. However,  in the warped ${\rm AdS}_3$ black hole for $\v>1$  the vanishing Dirichlet  boundary condition is not natural, since it has the finite effective potential at radial infinity, in contrast to the case for $\v=1$ \cite{bin, 0907}. Rather the Dirichlet boundary condition is not appropriate for the warped ${\rm AdS}_3$ black hole for $\v>1$, since it can have a non-vanishing wavefunction for the finite effective potential at radial infinity. 
Therefore, since the boundary condition at radial infinity is not obvious for $\v>1$ \cite{bin, bin2}, we need to find the correct quasinormal modes for $\v>1$ in a different way.  In obtaining quasinormal modes there are various methods. For example,  the quasinormal modes of  the BTZ black hole were obtained from the appropriate boundary condition at radial infinity in \cite{bir, cardo}.
However by using monodromy method the same quasinormal modes were also obtained in  \cite{0308, 0311, 0504}. In particular, without imposing any boundary condition at radial infinity the same quasinormal modes for the BTZ black hole were obtained in  \cite{0311, 0504}. Likewise, recently without imposing any boundary condition at radial infinity the quasinormal modes of the warped ${\rm AdS}_3$ black hole for $\v \ge 1$ for the massless scalar field were obtained by using monodromies at inner and outer horizons \cite{0912}. These are called  holographic quasinormal modes, which are given by
\begin{eqnarray} \label{hqh}
 \tilde \w_R^h  &\equiv & { {  \w_{R}^h   } \over l} = { { {-4  m } - i  (\v ^2+3) ({r_+} - {r_-})  k} \over  {2   {\left( \v ({r_+} + {r_-}) - \sqrt{(\v^2 +3) {r_+} {r_-} } \right) l} } }   ~, \\
 \label{hqh2}
 \tilde \w_L^h  & \equiv & { { \w_{L}^h   } \over l} = - i  { {(\v^2+3) k} \over {2 \v l} }~~,~~ (k \in N) ~,
\end{eqnarray}
where $m$ is the angular quantum number and negative imaginaries are taken for $e^{- i  \w t} $ of the wavefunction to have damped quasinormal modes. 
Note that $\tilde \w^h_{R(L)}$ is dimensionful and gives the transition frequency, $\tilde \w^h_{Rc(Lc)}$, which has the dimension of $E/\hbar$. 
When $\v=1$, the quasinormal modes are the same as what have been obtained  by imposing the vanishing Dirichlet  boundary condition at radial infinity in \cite{bin, bin2}, which is the appropriate boundary condition only for $\v=1$.  

At large $k$ for a fixed $\vert m \vert$, in particular for $k \gg \vert m \vert$,  the two possible transition frequencies are obtained from the two families of the quasinormal modes (\ref{hqh}) and (\ref{hqh2}) as follows:
\begin{eqnarray} \label{addqnm}
\tilde \w_{Rc}^h  &=& {  {\v ^2+3} \over {2 \v l}}   {  {\v  ({r_+} - {r_-})} \over  { \v  ({r_+} + {r_-}) - \sqrt{(\v^2+3) {r_+} {r_-} }  } }  \nonumber \\
&=&{{\left(\v^2 +3 \right)} \over {2 \v l} } {1\ \over{3 G \M}}   \sqrt{  { {6 G  \v} \over {5  \v^2 +3} }  \left(6  G  \M^2  \v - {\left(\v^2 +3\right)} {\J/ l} \right) }  ~, \\
 \label{addqnm2}
\tilde \w_{Lc}^h  &=& { {\v^2+3} \over {2 \v l} } ~,
\end{eqnarray}
where  (\ref{u1}) and (\ref{u2}) are used in rewriting in terms of $\M$ and $\J$.
From the formula (\ref{mf}) the action variable $\I$ is obtained and quantized;
\begin{equation} \label{hmff}
\I = \int { dE \over {\tilde \w_c ^h} } =  \int  { d\M \over {\tilde \w_c ^h} } = n  \hbar ~,
\end{equation}
where  the change of the energy of the warped ${\rm AdS}_3$ black hole is considered as the change of  the ADT mass $\M$.
Therefore two action variables $\I_R$ and $\I_L$ are obtained and quantized as follows:
\begin{eqnarray} \label{add3}
\I_{R} &=& \int { {d\M} \over {\tilde \w_{Rc}^h }} ={ {l \sqrt { \left( 5 {\v}^{2}+3 \right) \left( 6 G {\M}^{2} 
\v - \left( {\v}^{2}+3 \right)\J /l  \right) } }  \over { \sqrt{6  G \v} \left( {\v}^{2}+3
 \right) }}  = n_R  \hbar ~, \qquad \\
\label{add4}
\I_{L} &=& \int  { d\M \over {\tilde \w_{Lc}^h } } = {{2 \v  l} \over {\v^2 +3}}  \M= n_L  \hbar ~.
\end{eqnarray}
From these quantization conditions, we find the spectra of  $A_{tot}$ and $A_{sub}$  which are given by  (\ref{add}) and(\ref{add2});
\begin{eqnarray}
A_{tot} = { {96 \pi  G \v^2} \over {5 \v^2 +3} }  n_R  \hbar ~~,~~ A_{sub} = {24 \pi G}  n_L  \hbar ~.
\end{eqnarray}
The total horizon area  $A_{tot}$ is equally spaced and dependent on the coupling constant $\v$, and the area difference $A_{sub}$ is also equally spaced and independent of the coupling constant $\v$. Therefore the spectra of  the outer and inner horizon areas are obtained as follows:
\begin{eqnarray} \label{hqare}
A_{out} &=&  { {48 \pi  G   \v^2} \over {5 \v^2+3} }  \left(   n_R + {  {5 \v^2+3} \over {4 \v^2} }  n_L\right) \hbar ~, \\ \label{hqare2}
 A_{in} &=& \pm { {48 \pi  G   \v^2} \over {5 \v^2+3} }  \left(   n_R  - {  {5 \v^2+3} \over {4 \v^2} }  n_L\right) \hbar ~,
\end{eqnarray}
where $+$ sign in $ A_{in}$  with $ n_R > {  {5 \v^2+3} \over {4 \v^2} }  n_L $  corresponds to  the case (i) and   $-$ sign in $ A_{in}$ with $ n_R < {  {5 \v^2+3} \over {4 \v^2} }  n_L $  corresponds to the case (ii).
These area spectra are not equally spaced   and dependent on the coupling constant $\v$.
For the $\v=1$ case, the area spectra are equally spaced as follows:
\begin{equation} \label{60}
A_{out} = 6   \pi  G  (n_R + 2 n_L )   \hbar  ~~,~~ A_{in} = 6   \pi  G (n_R -2  n_L)   \hbar  ~,
\end{equation}
where $ n_R > 2 n_L  $.
So, the spacing of the area spectra is given by $\triangle A_{out} =\triangle A_{in}=6   \pi  G$. 
The area spectra (\ref{60}) are the exactly same as for the rotating BTZ black hole with the Chern-Simons term for $\v=1$ in the previous section. This is natural because the warped ${\rm AdS}_3$ black hole for $\v=1$ is the rotating BTZ black hole in a rotating frame \cite{strom}. 
Note that this spacing is different from the rotating BTZ black hole case without the Chern-Simons term, which is given by $\triangle A_{out} =\triangle A_{in}=8   \pi  G$ in \cite{sk}. So, it can be considered that the Chern-Simons term affects the spacing of the area spectra.

The entropy spectrum is obtained by 
rewriting the entropy expression (\ref{went})  in terms of the horizon areas as follows:
\begin{eqnarray} \label{add5}
S &=&  {\frac {\pi  l   \left( \left( 9 {\v}^{2} +3 \right) {r_+} - \left( {\v}^{2}+3 \right) {r_-} -4 \v\sqrt { ( {\v}^{2}+3)  {r_+} 
{r_-} }  \right) }{
24 G \v \hbar}} \nonumber \\
& =&  { {(9 \v^2 +3) A_{out} \pm (\v^2 +3) A_{in}} \over {48 G \v^2 \hbar}} ~,
\end{eqnarray}
where $+$ is for the case (i) and $-$ is for the case (ii).
Because of the Chern-Simons term, the entropy is proportional to not only the outer horizon area but also the inner horizon area, so that the Bekenstein-Hawking area law, $S={ A_{out} \over {4 G \hbar}}$, does not hold anymore.
%
Notice that the entropy in terms of the horizon areas is dependent on $\v$. However a remarkable cancellation of $\v$ dependence happens when we substitute (\ref{hqare}) and (\ref{hqare2}) into (\ref{add5}). We find that the entropy spectrum is equally spaced as follows:
\begin{eqnarray} \label{adden}
S  = 2 \pi   (n_R+n_L) ~.
\end{eqnarray}
This entropy spectrum is independent of the coupling constant $\v$ and the spacing  is given by $\triangle S= 2 \pi$.  The result (\ref{adden}) is the same for both cases (i) and (ii). 
Furthermore this is the exactly same as for the rotating BTZ black holes with and without the Chern-Simons term. 
It implies that the entropy spectra of the black holes have a universal behavior regardless of the presence of the gravitational  Chern-Simons term. Therefore  we would like to claim that the entropy spectrum rather than the area spectrum of a black hole should be equally spaced.

\section{Entropy spectrum as criteria for quasinormal modes}

In certain cases there are several different calculations for quasinormal modes of a black hole, with different results. As mentioned before, for example, there was another work on the quasinormal modes for the warped ${\rm AdS}_3$ black hole  \cite{bin, bin2}.  These quasinormal modes are obtained from the vanishing Dirichlet boundary condition which is not natural for the case $\v>1$. 
In this section we would like to show that the universality of the equally spaced entropy spectrum can be taken as  criteria for determining correct quasinormal modes. We propose that the quasinormal modes of a black hole which give rise to equal spacing in its entropy spectrum are the correct ones.
 
For that purpose we will show how the area and entropy spectra are given, when the quasinormal modes of the warped ${\rm AdS}_3$ black hole for $\v >1$ which are obtained from the vanishing Dirichlet  boundary condition are used.
From the quasinormal modes obtained in \cite{bin, bin2} the two transition frequencies are obtained as follows:
\begin{eqnarray}\label{w}
\fl  \tilde \w_{R c} \equiv { {  \w_{R c} } \over l}  &=& {\frac { ({\v}^{2}+3)  ( {r_+} - {r_-} )   }{2  \v   \left( {r_+} + {r_-} \right) -2 \sqrt { \left( {\v}^{2}+3 \right)  {r_+} {r_-}}-\sqrt {3 ({\v}^{2}-1 ) } \left( {r_+} - {r_-} \right) }}  {1 \over l} ~,\\
\label{w1}
\fl  \tilde \w_{L c} \equiv { {  \w_{L c} } \over l}  &=& {{ 2  \v+\sqrt{ 3 (\v^2 -1)} } \over l} ~. 
\end{eqnarray}
Note that $\tilde \w_{Rc(Lc)}$ is dimensionful and identified with the transition frequency which has the dimension of $E/\hbar$. 
When the formula (\ref{mf}) with $dE=d\M$ is applied with these two possible transition frequencies, we obtain the area spectra  as follows:
\begin{eqnarray} \label{was}
A_{out} &=& {{48  \pi  G  {\v}^{2}} \over{ 5 {\v}^{2}+3 }} \left( n_R+  {\frac {  16 {\v}^{3}+  \left( 9 {\v}^{2}+3 \right) \sqrt {3 ({\v}^{2}-1) }  }{2 \v
 \left( {\v}^{2}+3 \right) }}   n_L  \right) \hbar ~, \\
\label{was1}
A_{in} &=& \pm {{48 \pi G  {\v}^{2}} \over {  5 {\v}^{2}+3 }} \left( n_R -  {\frac {  4 \v+\sqrt {3 ({\v}^{2}-1)}    }{2 \v}} n_L \right) \hbar ~,
\end{eqnarray}
 where $+$ sign in $ A_{in}$  with $ n_R >  {\frac {  4 \v+\sqrt {3 ({\v}^{2}-1)}    }{2 \v}} n_L  $ corresponds to  the case (i)  and  $-$ sign in $ A_{in}$ with $ n_R <  {\frac {  4 \v +\sqrt {3 ({\v}^{2}-1)}    }{2 \v}} n_L  $ corresponds to  the case (ii).

Using these area quantizations, the entropy (\ref{add5})  are  quantized for both cases (i) and (ii) as follows:
\begin{equation} \label{ct13}
S   =    {2 \pi}    \left( n_R + { { {   7 {\v}^{2}-3+4 \v \sqrt {3 ({\v}^{2}-1)}  } \over {  {\v}^{2}+3   } } }   n_L  \right) ~.
\end{equation}
This is not equally spaced and depends on the coupling constant $\v$.
No cancellation of $\v$ dependence happens here.
There are  general arguments that the entropy should be equally spaced \cite{medved2, kotwal}.
We expect these types of arguments should go through even in the presence of the gravitational Chern-Simons term, and thus we expect equally spaced entropy spectrum. When a quasinormal mode fails to give equally spaced entropy, presumably it is an erroneous one.
%
Therefore in the sense that the entropy spectrum should have a universal behavior as an equally spaced one, the non-equally spaced entropy spectrum (\ref{ct13}) with $\v$ dependence implies that the quasinormal modes obtained  from the vanishing Dirichlet boundary condition at radial infinity in \cite{bin, bin2} are not correct  for $\v>1$. 
Therefore we conclude that the correct entropy spectra of the warped ${\rm AdS}_3$ black hole for $\v >1$  are given by  (\ref{adden}), and not by (\ref{ct13}).

\section{Conclusion}

We calculated the area and entropy spectra of the black holes in the topologically massive gravity. 
From the quasinormal modes of the black holes, we obtained  possible transition frequencies between quantum levels of the black holes.
%
By Bohr's correspondence principle, a quantum black hole with a transition frequency can be regarded as a classical system of periodic motion with the transition frequency as an oscillation frequency in the semiclassical limit. The action variable of the classical system is identified  and quantized via Bohr-Sommerfeld quantization. From this we obtained the area and entropy spectra of  the BTZ black hole and the warped ${\rm AdS}_3$ black hole under the consideration of the gravitational Chern-Simons term.

First we considered the BTZ black hole with the Chern-Simons term.
For the non-rotating case we found that the area and entropy spectra are equally spaced and the exactly same as for the non-rotating BTZ black hole without the Chern-Simons term obtained in \cite{sk}. This is because the Chern-Simons term does not affect the energy and entropy of the black hole.
%
For the rotating BTZ black hole with the Chern-Simons term  the spectra of the outer and inner horizon areas are not equally spaced   and dependent on the coupling constant $\v$. 
However the entropy spectrum is equally spaced and independent of the coupling constant $\v$.
Furthermore this  entropy spectrum is the exactly same as for the rotating BTZ black hole without the Chern-Simons term obtained in \cite{sk}. 
In particular for the rotating BTZ black hole with the Chern-Simons term for $\v =1/3$, which corresponds to chiral gravity, only the total horizon area spectrum was obtained and equally spaced. So, the spectra of the outer and inner horizon areas were not obtained. However we could find the  equally spaced entropy spectrum, since the entropy is only proportional to the total horizon area. 
The entropy spectrum has the same spacing as for the rotating BTZ black hole for other values of $\v$, i.e. $\triangle S=2 \pi$.

Next, the warped ${\rm AdS}_3$ black holes for  $\v \ge1$ were considered. 
 It is known that  the warped ${\rm AdS}_3$ black hole for $\v=1$ is related to the rotating BTZ black hole by a coordinate transformation.  In this sense we obtained the consistent result that  the area and entropy spectra of the warped ${\rm AdS}_3$ black hole for $\v=1$ are the exactly same as for the rotating BTZ black hole with the Chern-Simons term for $\v=1$. We found that  the both area and entropy spectra are equally spaced.
%
For the warped ${\rm AdS}_3$ black hole for  $\v >1$, the quasinormal modes were obtained by imposing the vanishing Dirichlet boundary condition at radial infinity \cite{bin, bin2}. But this boundary condition is not appropriate, since the effective potential in the wave equation is finite at radial infinity. 
There can be a  non-vanishing wavefunction at radial infinity. 
This is different from usual AdS black holes such as the BTZ black hole for which  the vanishing Dirichlet boundary condition  should be imposed at radial infinity because of  the divergent effective potential.
There is another calculation for the quasinormal modes of the warped ${\rm AdS}_3$ black hole \cite{0912}, obtained without imposing any boundary condition at radial infinity.
Using these quasinormal modes obtained in \cite{0912}, we found that the entropy spectrum is equally spaced and independent of the coupling constant $\v$, even though the area spectra are not equally spaced and dependent on the coupling constant $\v$. 
Moreover the entropy spectrum is the exactly same as for the rotating BTZ black holes with and without the Chern-Simons term. So, it implies that the entropy spectrum has a universal behavior.

This universality that the entropy spectrum of a black hole is equally spaced can be taken as criteria for determining correct quasinormal modes. We proposed that the correct quasinormal modes should result in the equally spaced entropy spectrum. 
%
By calculating the entropy spectrum with quasinormal modes in two different methods, we found two different entropy spectra. Of course only one of these should be correct. What we found is that  the quasinormal modes obtained from the vanishing Dirichlet boundary condition \cite{bin, bin2} does not give equally spaced entropy spectrum. 
On the other hand, the other quasinormal modes  obtained without imposing any boundary condition at radial infinity \cite{0912} does give equal spacing. Therefore the latter should be the correct one.

In this paper we extended the method for black hole quantization  proposed in the previous work \cite{sk} to the black holes in topologically massive gravity theory.
We found that the entropy spectra of the black holes are equally spaced, even though the area spectra  are not equally spaced.
Therefore, it should be considered that the entropy spectrum is more `fundamental' than area spectrum.
This has been observed before in  \cite{wei, kotwal}, where the entropy spectrum of the five dimensional black hole with the Gauss-Bonnet term is equally spaced, even though the area spectrum is not equally spaced.
%
 For the future investigation, it will be worth to find if entropy spectra for other black holes in higher dimensions and different gravity theory have universal behavior. For example, the quantization of the four dimensional black holes, in particular for the Kerr black hole whose area spectrum is not clear yet, as was explained in the introduction.
%
It is expected that the equally spaced entropy spectrum of a black hole can be used for various investigations on a quantum black hole as a quantum property of a black hole.
For example, very recently the equally spaced entropy spectrum was used as what is associated with the information of a black hole in the work on the emergent gravity  \cite{ver}.
%

\ack{ This work of YK and SN  was supported by the National Research Foundation of Korea(NRF) grant funded by the Korea government(MEST) (No. 2009-0063068)  and the work of YK was also supported  by the National Research Foundation of
Korea(NRF) grant funded by the Korea government(MEST) (No. 2009-0085995). YK thanks Dr. Jong-Dae~Park for useful discussions.}


\end{document}